\documentclass[useAMS,usenatbib]{mn2e}
\usepackage{graphicx}
\usepackage{array}
\usepackage{float}
\usepackage{hhline}

\title[Multi-mode pulsation of GD 154]{Multi-mode pulsation of the ZZ Ceti star GD 154}
\author[M. Papar\'o, Zs. Bogn\'ar, E. Plachy, L. Moln\'ar, P. A. Bradley]{M. Papar\'o$^{1}$\thanks{E-mail:
paparo@konkoly.hu}, 
Zs. Bogn\'ar$^{1}$, 
E. Plachy$^{2}$, 
L. Moln\'ar$^{1}$
and P. A. Bradley$^{3}$
\\ 
$^{1}$Konkoly Observatory, MTA CSFK, Konkoly Thege u. 15-17, H--1121 Budapest, Hungary \\
$^{2}$Department of Astronomy, E\"otv\"os University, P.O. Box 32., H--1518 Budapest, Hungary\\
$^{3}$XCP-6, MS F699, Los Alamos National Laboratory, Los Alamos, NM 87545, USA\\
}

\begin{document}

\date{Accepted 2012 ..... Received 2012 ...; in original form 2012 ...}

\pagerange{\pageref{firstpage}--\pageref{lastpage}} \pubyear{2012}

\maketitle

\label{firstpage}

%%%%%%%%%%%%%%%%%%%%%%%%%%%%%%

\begin{abstract}

We present the results of a comparative period search on different timescales and 
modeling of the ZZ Ceti (DAV) star GD 154. We determined six frequencies as normal modes and 
four rotational doublets around the ones having the largest amplitude. Two normal modes 
at 807.62 and 861.56\,$\mu$Hz have never been reported before. A rigorous test revealed 
remarkable intrinsic amplitude variability of frequencies at 839.14 and 861.56\,$\mu$Hz 
over a 50-day timescale. In addition, the multi-mode pulsation changed to monoperiodic pulsation with an 
843.15\,$\mu$Hz dominant frequency at the end of the observing run. The 2.76\,$\mu$Hz average 
rotational split detected led to a determination of a 2.1-day rotational period for GD 154. We searched for model 
solutions with effective temperatures and $\rmn{log\,} g$ close to the spectroscopically determined ones. 
The best-fitting models
resulting from the grid search have $M_\rmn{H}$ between $6.3*10^{-5}$ and $6.3*10^{-7} M_*$, which means 
thicker hydrogen layer than the previous studies suggested. 
Our investigations show
that mode trapping does not necessarily operate in all of the observed
modes and the best candidate for a 
trapped mode is at $2484\,\mu$Hz. 
\end{abstract}

\begin{keywords}
techniques: photometric --
stars: individual: GD 154 --
stars: interiors --
stars: oscillations --
white dwarfs.
\end{keywords}

%%%%%%%%%%%%%%%%%%%%%%%%%%%%%%

\section{Introduction}

It has been well established that \textit{g}-mode white dwarf 
pulsators show diversity in light variation from the simple sinusoidal to the mightily complicated ones. 
The latter variation accompanied by nonlinear features due to harmonics and combination 
frequencies (see e.g. the review of \citealt{a3}). 

It has also been pointed out that pulsational amplitudes and/or the frequency content 
of light curves might vary on short timescales. These phenomena are quite common 
amongst the cooler DAVs and DBVs, and also the PNNV stars \citep{a4}. The amplitude 
variability can be helpful to asteroseismology as different normal modes become 
detectable over time and therefore the number of known modes increase (see e.g. the 
case of G29-38, \citealt{a7}). However, amplitude variations on timescales of 
weeks and months can make the accurate determination of individual modes difficult 
or even impossible using the standard Fourier deconvolution technique. These 
variations occur sometimes as sudden effects as in the case of the `everchanging' 
GD 358, the most spectacular representative of the phenomenon \citep{a5,a6}. In 
spite of intense observational efforts, the origin of these phenomena still remains unknown. A 
possible explanation, which can be tested relatively easily, is mode beating 
due to different unresolved pulsational modes (such as rotational splitting). 

The DAVs lying near the red edge of the ZZ Ceti instability strip (GD 154 is an example) 
are characterized by long periods and complex pulsational behaviour. 
Changes in pulsation from a single mode status (connected by harmonics and not trivially by subharmonics) 
to a multi-mode status have been seen for GD 154 during the previous observations. 
Since the seemingly new appearance of a different normal mode can be caused only by the 
amplitude variation of the ever excited normal modes, we decided to follow the pulsational 
behaviour for an observational season. We present our comparative analyses on different 
timescales. It was the increased number of pulsational periods determined by our dataset that allowed us to invstigate the star from an asteroseismological point of view. 
We also discuss the related results 
in the paper.

\section[]{Observations and data reduction}

We used the 1-m Ritchey-Chr\'etien-Coud\'e telescope at Piszk\'estet\H{o} mountain 
station of Konkoly Observatory for data collection on GD 154. The observations were 
made with a Princeton Instruments VersArray:1300B back-illuminated CCD camera without 
any filter. Our observing season spanned six months in 2006. Altogether, 90 hours of 
photometric data were collected on 19 nights. 30\,s integration times were used. The 
longest continuous single-night light curve was more than 9-hour long. The shortest 
one that we found useful to determine the frequency content covers two full cycles. 
Details of the observations are provided in Table~\ref{tabl:log}.

\begin{table}

\caption{Log of observations of GD 154. Five subsets were created based on the 
closely spaced nights.}
\label{tabl:log}

\begin{center}
\begin{tabular}{p{4mm}p{6mm}rcrr}
\hline
Run & Subset & UT date & Start time & Points & Length\\
No. & No. & (2006) & (BJD-2\,450\,000) & & (h)\\
\hline
01 & 1 & Feb 03 & 3769.520 & 372 & 4.04\\
02 & 1 & Feb 05 & 3771.575 & 335 & 3.28\\
03 & 1 & Feb 07 & 3773.519 & 340 & 4.06\\
04 & 2 & Mar 02 & 3797.430 & 585 & 5.66\\
05 & 2 & Mar 06 & 3801.481 & 484 & 4.72\\
06 & 2 & Mar 07 & 3802.285 & 787 & 9.11\\
07 & 2 & Mar 08 & 3803.305 & 684 & 6.60\\
08 & 3 & Mar 31 & 3826.364 & 659 & 6.60\\
09 & 3 & Apr 02 & 3828.288 & 291 & 5.03\\
10 & 3 & Apr 04 & 3830.357 & 451 & 4.42\\
11 & 4 & Apr 21 & 3847.385 & 493 & 5.45\\
12 & 4 & Apr 22 & 3848.321 & 709 & 6.88\\
13 & 4 & Apr 23 & 3849.316 & 549 & 6.85\\
14 & 4 & Apr 24 & 3850.380 & 468 & 5.20\\
15 & 4 & Apr 25 & 3851.332 & 541 & 6.44\\
16 & 5 & Jul 13 & 3930.340 & 65 & 0.71\\
17 & 5 & Jul 17 & 3934.335 & 92 & 1.06\\
18 & 5 & Jul 18 & 3935.336 & 161 & 1.83\\
19 & 5 & Jul 19 & 3936.342 & 150 & 1.65\\
Total: & & & & 8216 & 90.19\\
\hline
\end{tabular}
\end{center}
\end{table}

\begin{figure}
\begin{center}
\includegraphics[width=6cm]{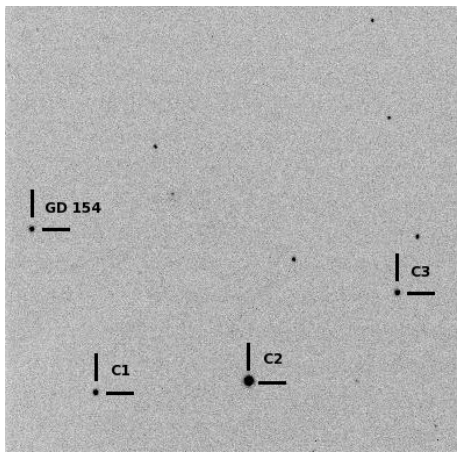}
\end{center}
\caption{The CCD field with the variable and the comparison stars. The marked 
stars C1, C2, C3 were used to construct a reference light curve.}
\label{fig:map}
\end{figure}

\begin{figure}
\begin{center}
\includegraphics[width=8.3cm]{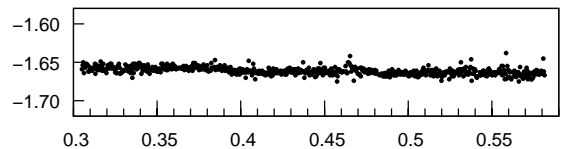}
\end{center}
\caption{Differential light curve of C1 to the average of the three reference 
stars is shown (C1, C2, C3), obtained on JD\,2\,453\,803. The standard deviation is 
0.003\,mag.}
\label{fig:lccomp}
\end{figure}

The original images were reduced using standard \textsc{iraf}\footnote{\textsc{iraf} 
is distributed by the National Optical Astronomy Observatories, which are operated 
by the Association of Universities for Research in Astronomy, Inc., under cooperative 
agreement with the National Science Foundation.} routines. Bias, dark and flat corrections 
were applied on the frames. We performed aperture photometry using the \textsc{iraf daophot} 
package, setting the aperture size to two times the average FWHM on the given night, and 
applied differential photometry on the stars. All the time data were calculated to 
Baricentric Julian Date (BJD).

We obtained $BVRI$ photometry and checked the $B-V$ indices of the potential comparison 
stars on the frames, but we found that they were rather different than the value of GD 154 
($B-V=$ 0.18). As there was no candidate for a comparison star with similar colour to our 
target, we tested all the stars on our CCD field by computing the average magnitudes to 
construct a reference light curve. We found that the most constant signal with the lowest 
standard deviation value could be constructed by averaging the data of the 
three brightest stars around the variable (see Fig.~\ref{fig:map}). A part of a differential 
light curve of the reference system and a check star's (C1) averaged data is presented in 
Fig.~\ref{fig:lccomp}. We applied the widely used polynomial fitting method to remove the 
effect of the atmospheric extinction. The low-frequency part of the Fourier Transform (FT) 
is biased by this filtering (up to 20-30\,cd$^{-1}$), for which reason we did not focus on low 
frequencies in our Fourier analysis.

\section[]{Frequencies in GD 154}

\begin{figure}
\begin{center}
\includegraphics[angle=0,width=8.6cm]{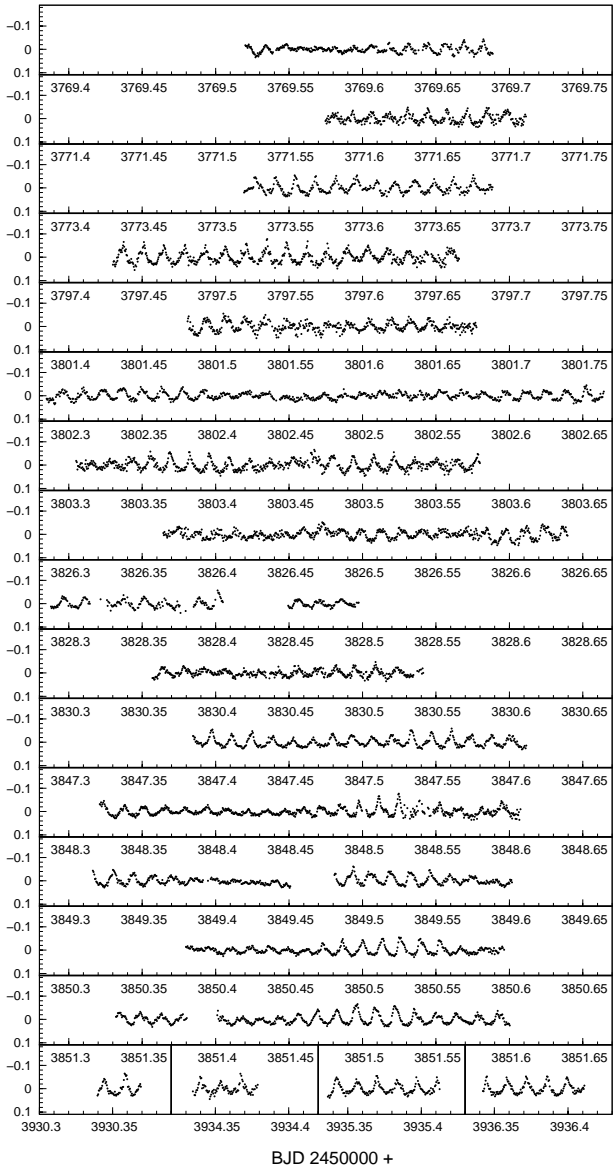}
\end{center}
\caption{Normalized differential light curves of the 19-night run on GD 154.}
\label{fig:lc}
\end{figure}

The light curves of GD 154 during 2006 can be seen in Fig.~\ref{fig:lc}. The different 
minimum to maximum amplitude of the cycles suggest an overall multi-mode behaviour, except 
for the regular cycles of the last nights.
To get an overall view of 
the pulsation modes of GD 154, we present here the modes determined in different epochs.

\subsection[]{Frequencies in the previous data}

The light variation of GD 154 was discovered in 1977 by \citet{a1}. The star 
showed nonlinear monoperiodic pulsation with F = 843\,$\mu$Hz main frequency, 
harmonics (2F, 3F, 4F, 5F) and `intermediate frequencies' (1.52F, 2.53F, 3.53F). The
 pulsation was regular except on the last night of the observations, when the 
light curve changed dramatically as the 1.52F mode became dominant. The authors 
suggested that the energy exchanged between the two modes due to weak non-linearity.

The existence of the near half-integer frequencies in the power spectra is not
unique. Similar pulsational behaviour was observed in PG 1351+489
(DBV, \citealt{a12}) and in G191-16 (DAV, \citealt{a13}).  
According to a possible explanation an independent
 \textit{g}-mode can be excited near 1.5F, as we see in the interpretation of
the data obtained on BPM 31594 \citep{a15}.

The Fourier spectral features and the sudden change in the dominant period made
GD 154 interesting enough to be the target of the Whole Earth Telescope (WET, 
\citealt{nather1}) campaign in 1991. The 12-day quasi-continuous observations and 
the follow-up campaign organized a month later resulted in another interpretation of
 the pulsation. Three independent modes ($f_1 =$ 842.8\,$\mu$Hz, $f_2 =$ 918.6\,$\mu$Hz, 
$f_3 =$ 2484.1\,$\mu$Hz) and their triplet components were found in the
 dataset. All the other peaks in the power spectra were explained
as linear combinations and harmonics. Non-linear behaviour was clearly confirmed 
but no drastic period change was observed. Surprisingly, they did not find peaks
 around half-integer frequencies \citep{a2}.

In 2004, an analysis of a two-site observational campaign \citep{a14} reported changes
in the frequency and amplitude content. Some additional frequencies appeared, 
however none was near the subharmonic. They were interpreted as new excited modes
at 786.5\,$\mu$Hz, 885.4\,$\mu$Hz and 1677.7\,$\mu$Hz. However, the authors did not
 find the frequencies at 842.8\,$\mu$Hz and 918.6\,$\mu$Hz that were observed before.

We can summarize that although four modes (786.5, 842.8, 885.4 and 918.6\,$\mu$Hz) 
were reported in a 
132\,$\mu$Hz range, not more than two were present during a
given observing run. None of them was present in every dataset, although one of 
them ($\approx$843\,$\mu$Hz) was present in 1977 \citep{a1} and 
in 1991 (WET campaign, \citealt{a2}) with nearly the same value. The different
solutions raise the possibility that (1) different modes are excited from time
to time or (2) their amplitude is changing to below or above the detection limit, or 
(3) the actual frequency content interfering with the alias patterns of a given data
 distribution results in a seemingly changeable frequency content.

\subsection[]{Analyses of the new data}

A standard frequency analysis was performed with the Multi-Frequency Analyzer (MuFrAn) 
time-string tool \citep{kollath1, csubry1}. The software package is a collection of 
methods for period determination, sine-wave fitting for observational data and graphic 
routines for visualization of the results. In each step, a fine tuning of the actual multi-frequency solution was carried out to avoid the disadvantage of prewhitening. At the same time, both the amplitudes and the fit 
were checked to avoid any misidentification of a peak as a real frequency. Investigations 
for significance and errors were obtained with the Period04 package \citep{lenz1}.

\begin{figure}
\begin{center}
 \includegraphics[width=9cm]{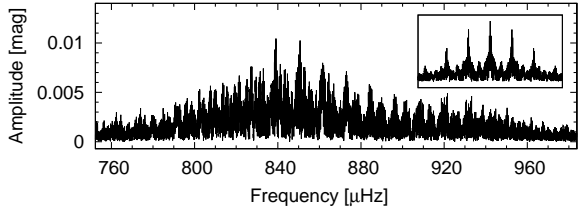}
\end{center}
 \caption{Fourier spectrum of the whole dataset. The window function is given in the insert.}
\label{fig:wholeft}
\end{figure}

The Fourier spectrum of our whole dataset (over 167 days) revealed the complexity of the 
resolution based on a set of single site observations (Fig.~\ref{fig:wholeft}). 

The overall appearance, a pyramid-like feature, suggests that more frequencies are present 
in the 760 -- 960\,$\mu$Hz frequency range. The 
fine structure is more complicated than the spectral window (insert in Fig.~\ref{fig:wholeft}) and suggests that 
the amplitude of the modes may have changed during the whole timebase.

We carried out comparative analyses on different timescales (weeks and a whole observational season)  
that resulted in different spectral window patterns. 
On a medium-size timescale (regularity of telescope allocation)  
the frequency resolution is acceptable, and the amplitude variation is still not masked. The baseline of 
the whole dataset gives a better resolution, but the amplitudes are not correct when 
amplitude variation is going on in a shorter timescale.

\begin{table}
  \caption{The frequency content of the whole data and two subsets}.
  \label{tabl:w2nd4th}
  \begin{center}
\begin{tabular}[h!]{rrrrrrr}
\hline
\multicolumn{1}{r}{Mode} & \multicolumn{2}{c}{Whole data} & \multicolumn{2}{c}{2nd subset} & \multicolumn{2}{c}{4th subset}\\
& \multicolumn{1}{c}{Freq.} & \multicolumn{1}{c}{Ampl.} & \multicolumn{1}{c}{Freq.} & \multicolumn{1}{c}{Ampl.} & \multicolumn{1}{c}{Freq.} & \multicolumn{1}{c}{Ampl.}  \\
\multicolumn{1}{r}{No.} &  \multicolumn{1}{c}{$\mu$Hz} & \multicolumn{1}{c}{mmag} & \multicolumn{1}{c}{$\mu$Hz} & \multicolumn{1}{c}{mmag} & \multicolumn{1}{c}{$\mu$Hz} & \multicolumn{1}{c}{mmag}  \\
 \hline
1 & 839.14  &  9.16   & 838.92  & 16.19  & 838.53 & 6.00   \\
2 & 843.15  &  9.46   &  -      & -      & 843.50 & 4.07   \\
3 & 844.65  &  7.22   & 845.17  & 3.82   & -      & -     \\
4 & 861.56  &  7.31   & -       & -      & 861.52 & 13.04  \\
5 & 864.27  &  4.23   & -       & -      & 863.98 & 4.39  \\
6 & 857.84  &  4.37   & -       & -      & -      & -      \\
7 & 807.62  &  3.99   & -       & -      & -      & -      \\
8 & 802.94  &  3.97   & -       & -      & 802.93 & 5.81   \\ 
9 & 809.20  &  3.76   & 809.77  & 8.33   & 810.17 & 3.18   \\
10 & 918.72 &  4.78   & 918.95  & 5.93   & 917.80 & 5.26   \\
11 & 920.41 &  4.11   & -       & -      & 920.24 & 7.10   \\
12 & 883.56 &  4.32   & 888.06  & 2.80   & 883.48 & 4.44   \\  
\hline
\end{tabular}
  \end{center}
\end{table}

Concerning the 760 -- 960\,$\mu$Hz frequency region, twelve
frequencies were identified in the {\it whole dataset}, and are given in 
Table~\ref{tabl:w2nd4th} (col. 2 and 3). The 0.035 $\mu$Hz Rayleigh frequency of the whole dataset guaranteed that the frequencies are properly resolved. According to the frequency differences three triplets, a doublet and a single frequency were found. 
We confirmed the frequencies found previously, except the
one at 786.45\,$\mu$Hz given by \citet{a14}. Our new discoveries are two modes 
at 807.62 and 861.56\,$\mu$Hz values. The increased number of normal modes puts more constrains for modelling.

\begin{figure}
\begin{center}
 \includegraphics[width=8cm]{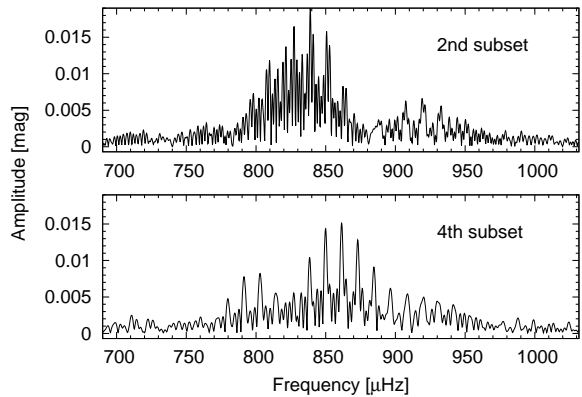}
\end{center}
 \caption{Comparison of the frequency content of the 2nd and 4th subsets. The frequency 
 content of GD 154 and especially the amplitudes changed from 3803 to 3847 BJD.}
\label{fig:2nd4th}
\end{figure}

Checking the time dependence of the frequency content, five subsets were created, but only the longest and the 
observationally most populated subsets' Fourier spectra (2nd and 4th) are presented 
here (Fig.~\ref{fig:2nd4th}). In the 2nd subset (four nights between 3797 and 3803 BJD) 
26.1 hours of observations were collected over 5.9 days. In the 4th subset (five nights 
between 3847 and 3851 BJD) more measurements were obtained (30.8 hours) but over a 
shorter timebase (3.9 days). The corresponding Rayleigh frequencies on the timebase of 
the subsets are 
1\,$\mu$Hz and 1.57\,$\mu$Hz. 
The alias pattern of the 4th 
subset is much cleaner due to the consecutive nights.

At first sight the Fourier spectrum of the 2nd and 4th subsets (upper and lower panels of 
Fig.~\ref{fig:2nd4th}) suggests that the frequency content of GD 154 and especially the 
amplitudes changed from 3803 to 3847 BJD.

We clearly recognize distinct peaks at 839\,$\mu$Hz  and 918\,$\mu$Hz in the 2nd subset and 
at 861\,$\mu$Hz and 802\,$\mu$Hz in the 4th subset. The accepted frequency content of the two 
subsets as a result of comparative analyses are given in Table~\ref{tabl:w2nd4th}.  
We found a doublet around the dominant mode in the 2nd subset and 
four doublets in the 4th subset, for a total of 5 and 9 frequencies, respectively. The triplets are not resolved on the subsets.  
Although the frequency values obtained on the different datasets agreed within 1\,$\mu$Hz in most cases, the amplitudes are remarkably different at different epochs.

\subsubsection[]{Results on data simulation}
\label{sect:tests}

Two points were checked in synthetic data. The frequency spacings of 839.14\,$\mu$Hz and 883.56\,$\mu$Hz frequencies to the newly found mode at 861.56\,$\mu$Hz are 22.42\,$\mu$Hz and 22.0\,$\mu$Hz, about twice the daily alias pattern (11.57\,$\mu$Hz), that could be confusing in the case of a single site observation. The reason of the apparently variable amplitudes of the 839.14\,$\mu$Hz and 861.56\,$\mu$Hz frequencies in the two subsets were also checked, whether they are interaction of the unresolved triplets or they represent intrinsic amplitude variation of normal modes.

Synthetic data were generated with the different combination of the possible normal modes, doublets and triplets, with constant amplitude for the time series of the whole season and the two subsets. The analyses of the synthetic data were conclusive.

The five possible normal modes can be obtained with high precision not only for the 
whole dataset but for the subsets, too. 
We confirmed that the new frequency at 861.56\,$\mu$Hz is a real normal mode, it cannot appear by the daily alias interaction. 
The interaction of doublets does not cause large amplitude changes in the subsets. 
The interaction of the three triplets on shorter timebase
can cause as large amplitude increase as the one we found in the analyses of our real 
time series. However, the amplitude ratio of the 838.92\,$\mu$Hz and 861.52\,$\mu$Hz frequencies did not change in the synthetic data.  
The amplitude of 838.92\,$\mu$Hz frequency was always higher in both subsets than the amplitude of
the 861.56\,$\mu$Hz frequency. 
We confirmed that both modes reveal real amplitude variations over a 50-day timescale. 
As we require a stable frequency solution for our single site
measurements, we present a solution that includes only the doublets.

\subsubsection[]{Harmonics and subharmonics}

\begin{figure}
\begin{center}
 \includegraphics[width=6.5cm]{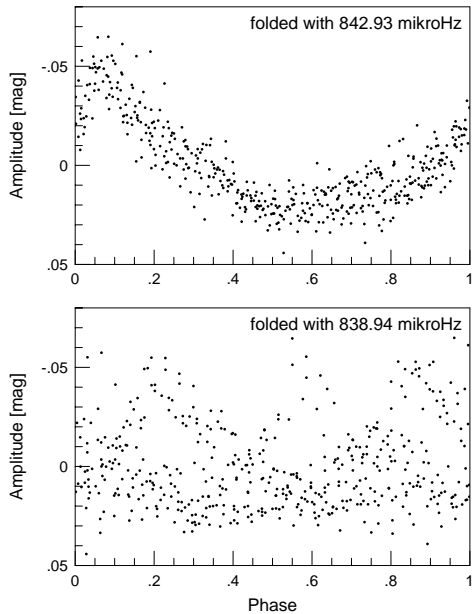}
\end{center}
 \caption{The light curves of the 5th subset (BJD\,3930 -- 3936) folded by the dominant mode 
 of the subset (upper panel) and by the dominant mode of the whole dataset (lower panel). The 
 regular arrangement of the upper panel suggests a monoperiodic pulsational phase of GD 154 
 during the given interval.}  
\label{fig:fig6}
\end{figure}

We performed frequency analyses in the region of harmonics and subharmonics, too. An independent frequency at 2484.14\,$\mu$Hz was definitely found on those nights when the star showed multimode pulsation. 

We found groups of peaks around 1652.78, 1711.81, 1752.31 and 1836.81\,$\mu$Hz  values but  
they do not correspond exactly to twice the value of the mother frequencies and it is hard 
to explain them as linear combinations. 
However, the second harmonic of the dominant mode (2517.48\,$\mu$Hz, the mother 
frequency is 839.14\,$\mu$Hz), near to the high frequency independent mode (2484.14\,$\mu$Hz), was clearly recognized. 

Linear combinations of the dominant mode and its doublet with the high frequency independent mode were detected at 3253.47 and 4155.09\,$\mu$Hz.  
Although these values are near the 4F and 5F 
values, they agree with the linear combination explanation much better. 

On the last, but unfortunately the shortest 
nights only a single mode at 842.93\,$\mu$Hz seems to be 
excited, instead of a multi-mode behaviour. It corresponds to the rotational component (at 843.15\,$\mu$Hz) that was dominant in the discovery and the WET runs. In Fig.~\ref{fig:fig6} we present the light curves 
of the 5th subset (BJD\,3930 -- 3936) folded by the dominant mode of the subset, 842.93\,$\mu$Hz 
(upper panel) and by the actual value of the dominant mode of the whole dataset, 838.94\,$\mu$Hz 
(lower panel). The narrow spread of the measurements in the upper panel proves that the star has a 
single mode dominant and this dominant mode is 842.93\,$\mu$Hz in the BJD\,3930 -- 3936 interval.
Unfortunately, we could not follow in details this status of the pulsation that seems to appear 
from time to time between the multi-mode states. On 20 and 21 April, 2007 GD 154 showed a 
pure monoperiodic pulsation state (Montgomery, private communication) with harmonics and subharmonics. 
The convective response timescale of the DAV star EC14012-1446 was compared partly to that value of GD 154 
obtained from the monoperiodic light curves \citep{a8}. 
Weak signs of 
the subharmonics (at 1195.60 and 1291.67\,$\mu$Hz) appeared mostly on our last short nights 
connected to the single mode pulsation. There were nights when the sign of the 2.5 times 
(2104.17\,$\mu$Hz) or 3.5 times (3003.47\,$\mu$Hz) values of the mother frequency could also  
be found.

\begin{table}
\caption{ The frequency content of GD 154 at 2006. Four doublets and two independent modes were found. 
The modes at 861.56 and 807.62 $\mu$Hz have never been reported before.}
\label{tabl:final}
\begin{center}
\begin{tabular}{rrrr}
\hline
Freq. & Period & Ampl. & Phase\\
($\mu$Hz) & (s) & (mmag) & (degree) \\
\hline
839.14 & 1191.7 & 8.38 & 314.4 \\
843.15 & 1186.0 & 6.90 & 1.9 \\
861.56 & 1160.7 & 5.76 & 276.5 \\
864.55 & 1156.7 & 4.91 & 152.8 \\
918.70 & 1088.5 & 4.44 & 215.6 \\
921.61 & 1085.1 & 3.88 & 94.6 \\
807.62 & 1238.2 & 4.64 & 218.9 \\
803.74 & 1244.2 & 4.38 & 144.0 \\
883.56 & 1131.8 & 4.12 & 337.4 \\
2484.14 & 402.6 & 3.50 & 82.6 \\
\hline
\end{tabular}
\end{center}
\end{table}

The Fourier parameters 
of the final accepted frequency content of GD 154 are given in Table~\ref{tabl:final}.

\subsubsection[]{Characteristic features in the light curves}

\begin{figure}
\begin{center}
 \includegraphics[width=7.5cm]{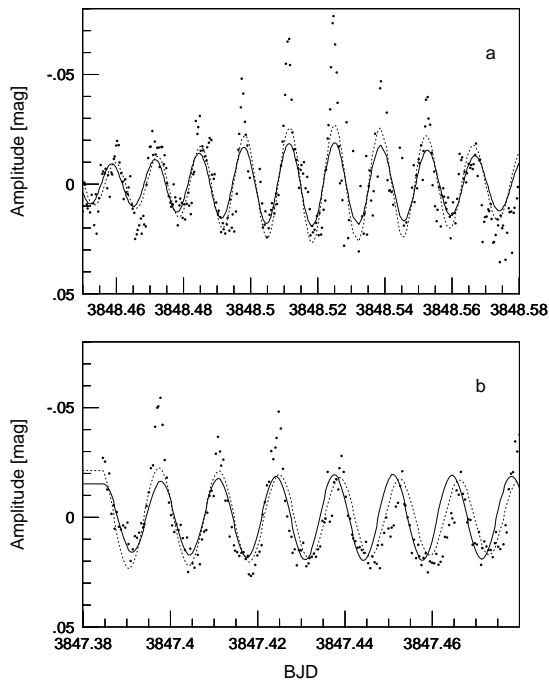}
\end{center}
 \caption{ Special features in the light curves of GD 154, $a$: intervals containing 4-6 cycles 
 where the measurements represent much higher amplitudes for the cycles than the fit (maybe 
 interaction of the pulsation with convection), $b$: alternating high and low amplitude cycles 
 follow each other (maybe chaotic behaviour of the pulsation). Fits for the solution for 
 the whole dataset (continuous line) and for the 4th subset (dotted line) are also given}
\label{fig:fig7}
\end{figure}

A fit, generated by the Fourier parameters given in Table~\ref{tabl:final}, describes the
general features of the light curve of GD 154. However, there are intervals containing some 
(4-6) cycles where the measurements represent much higher amplitudes for the cycles than the 
fit. Two special features were isolated. In panel \textit{a} of Fig.~\ref{fig:fig7}, the 
envelope of the measurements resembles a Gaussian profile with steep increase before the 
maximum and a steep decrease after it. The fit by our mode decomposition gives a much lower 
regular amplitude of the cycles. Both the solution (only the modes in the 760 -- 960\,$\mu$Hz 
interval) for the whole dataset (continuous line) 
and the separate solution for the 4th subset (dotted line) are compared. The latter slightly 
increased the amplitude of the fit but in the maximum-amplitude cycle only about 50\% of the 
amplitude is covered by the fit. Similar high-amplitude intervals are between BJD\,3803.34 -- 3803.44 
and 3850.46 -- 3850.56. The inclusion of the rotationally split components helped to increase the 
amplitude but it was still not enough to match the observed amplitude. It is hard to imagine such a 
missing pulsation mode that could help to fit the high-amplitude cycles without destroying the acceptable 
fit of the low-amplitude cycles. It is more probable that some additional physical process is superimposed 
on the pulsation creating high-amplitude phases. Maybe we are faced with an extra effect of 
convection with the pulsation.

A second characteristic feature is presented in panel \textit{b} of Fig.~\ref{fig:fig7}: alternating 
high- and low-amplitude cycles follow each other. Neither the frequency solution of the whole dataset, 
nor the 4th subset can fit the light curve of these cycles. Similar features can be found between 
BJD\,3847.51 -- 3847.56, 3851.48 -- 3851.54 and 3935.34 -- 3935.39. The presented case in panel 
\textit{b} shows a phase shift between the solution for the whole dateset and the 4th subset. The 
alternating high and low amplitude cycles remind us of the chaotic behaviour of the stellar pulsation.

\subsection{Rotational splitting}

\begin{table*}
\caption{We varied the stellar parameters and built model grids according to the minima and 
maxima values and step sizes given in the table.}
\label{tabl:grid}
\begin{center}
\begin{tabular}[!ht]{lcccccc}
\hline
Grid & $T_{\rmn{eff}}$ (K) & $M_*$ ($M_{\sun}$)& -log\,$M_{\rmn{He}}$ & -log\,$M_{\rmn{H}}$ & $X_{\rmn{O}}$ & $X_{\rmn{fm}}$\\
\hline
1 & 10\,600\,--\,11\,800 & 0.600\,--\,0.800\, & 2 & 4\,--\,11 & 0.5\,--\,0.9 & 0.1\,--\,0.5\\
2 & 10\,600\,--\,11\,800 & 0.600\,--\,0.800 & 2\,--\,3.5 & 4\,--\,11 & \multicolumn{2}{c}{core profiles by \citet{salaris1}}\\
Step sizes: & 200 & 0.005 & 0.5 & 0.2 & 0.1 & 0.1\\
\hline
\end{tabular}
\end{center}
\end{table*}

Thanks to regular observations over the whole season, we could find not only the normal modes but 
members of rotationally split frequencies. The direct determinations of the rotational splittings are 
4.01, 2.99, 2.91 and 3.88\,$\mu$Hz for the doublets presented in Table~\ref{tabl:final}, respectively.
Considering the three triplets listed in Table~\ref{tabl:w2nd4th}, we recognize that they are 
asymmetrically spaced and the $m=0$ to $+1$ splits are always smaller. This suggests that the 
$\approx920\,\mu$Hz doublet peaks, as also having small spacing, belong to $m=0$ and $m=+1$ modes.     

We use the term `triplets' for the multiplets we found, but without any mode identification, we cannot
say which ones are real triplets (dipole modes) or three of five possible components of quadrupole modes. The
asymmetric structure of these closely spaced modes also makes difficult to distinguish between
the $l=1$ and $l=2$ ones. Assuming that the triplet components at the dominant mode are high-overtone 
($k\gg1$) $l=1$ ones, calculating from the $2.76\,\mu$Hz average split value, the rotation period of the 
star is 2.1\,d, assuming solid body rotation.     

\citet{a2} determined $P_{\rmn{rot}} = 2.3\pm0.3$\,d, also derived from an asymmetric triplet with 
$\langle\delta f\rangle = 2.5\,\mu$Hz. This rotation period agrees with our result within $1\sigma$. 
\citet{a14} found $\approx$3\,$\mu$Hz frequency 
splittings ($\langle\delta f\rangle = 3.27\,\mu$Hz), corresponding to $P_{\rmn{rot}}=1.8$\,d. 
Regarding that at each epoch different frequency content were determined, the similar frequency 
spacings of the multiplets suggest they really correspond to stellar rotation at a constant rate.

Beside the direct determinations, we searched for characteristic frequency spacing values applying 
a more sophisticated method. We selected the five main frequencies in the 65 -- 85\,cd$^{-1}$ range 
and -- during a pre-whitening process -- the highest amplitude ones in their vicinity. Then we 
performed the Fourier analysis of the \textit{frequencies} obtained this way. That is we searched 
for regular spacing value(s) between the frequencies which may correspond to the rotational splitting 
phenomenon. To check our findings, we analysed not only the whole frequency list, but some subsets of 
the frequencies, too. Our results suggest characteristic spacing values being around 2.6 and 
3.7\,$\mu$Hz \citep{bognar1}. These values are also close to the directly determined and the 
previously detected ones.   

\section[]{Investigation of the main stellar parameters}

The efforts to determine the precise frequencies of the independent modes were aimed at 
investigating the interior structure of the star. Asteroseismology gives us the opportunity to provide 
constraints on the structure of the core, the hydrogen/helium layers, the mass of 
the star and to estimate the star's distance. For 
this purpose, we ran the White Dwarf Evolution Code (WDEC) originally written by Martin 
Schwarzschild and modified by \citet{kutter1}, \citet{lamb1}, \citet{winget1}, \citet{kawaler1}, 
\citet{wood1}, \citet{bradley1}, \citet{montgomery1} and \citet{kim1}.

The WDEC evolves a hot ($\sim$100\,000\,K) polytrope starter model down to the temperature we 
require, and gives an equilibrium, thermally relaxed solution to the stellar structure. For this 
model, the possible pulsation periods of $m=0$ are determined by solving the non-radial, adiabatic 
stellar pulsation equations \citep{unno1}. \citet{metcalfe1} created an integrated form of the 
evolution/pulsation codes, which allow us to obtain the period values with only one command. This 
way we can build model grids consisting of thousands of models in a very efficient way.  

We used the equation-of-state (EOS) tables of \citet{lamb2} in the core, the EOS tables of 
\citet{saumon1} in the envelope of the star, OPAL opacities updated by \citet{iglesias1} and the conductive opacities by \citet{itoh1, itoh2}. 
The WDEC treats the convection by means of the mixing length theory (MLT) of \citet{bohm1} using 
the $\alpha = 0.6$ parametrization according to the model calculations of \citet{bergeron2}. 
The hydrogen/helium transition zone was treated by equilibrium diffusion calculations, while the 
helium/carbon transition layer was parametrized.

We fitted the observed pulsation periods with the calculated dipole and quadrupole ones and 
searched for the best-fitting models. The goodness of the fit was described by the \emph{r.m.s.} 
value calculated by the following way:

%\[
\begin{equation}
\sigma_{r.m.s.} = \sqrt{\frac{\sum_{i=1}^{N} (P_i^{\rmn{calc}} - P_i^{\rmn{obs}})^2}{N}}
\label{equ1}
%\]
\end{equation}

\noindent where \textit{N} is the number of observed periods. The $\sigma_{r.m.s.}$ values were 
calculated using the \textsc{fitper} program of \citet{kim2}.

We built two model grids using different core composition profiles. In one case, we varied 
five input parameters of the WDEC: $T_{\rmn{eff}}$, $M_*$, $M_{\rmn{H}}$, $X_\rmn{O}$ 
(central oxygen abundance) and $X_{\rmn{fm}}$ (the fractional mass point where the oxygen 
abundance starts dropping) and fixed the mass of the helium layer at the $10^{-2}\,M_*$ `canonical' 
value. In the second scan, we varied the mass of the helium layer, but used the core profiles of 
\citet{salaris1} based on evolutionary calculations. For this second grid, only four stellar parameters 
were scanned. Table~\ref{tabl:grid} shows the parameter space covered by our grids and the step 
sizes applied. 

\begin{figure*}
%\begin{center}
  %\label{fig:params1}
  \includegraphics[width=8.7cm]{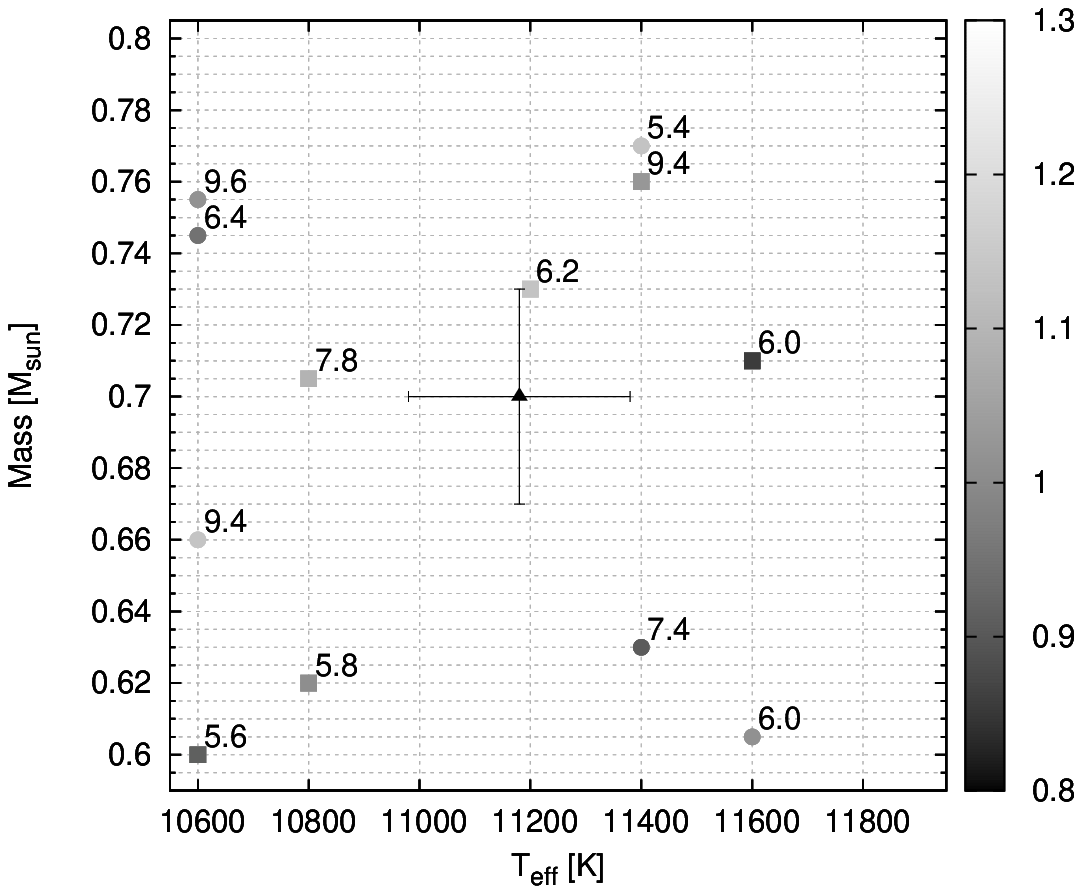}
  %\label{fig:params2}
  \includegraphics[width=8.7cm]{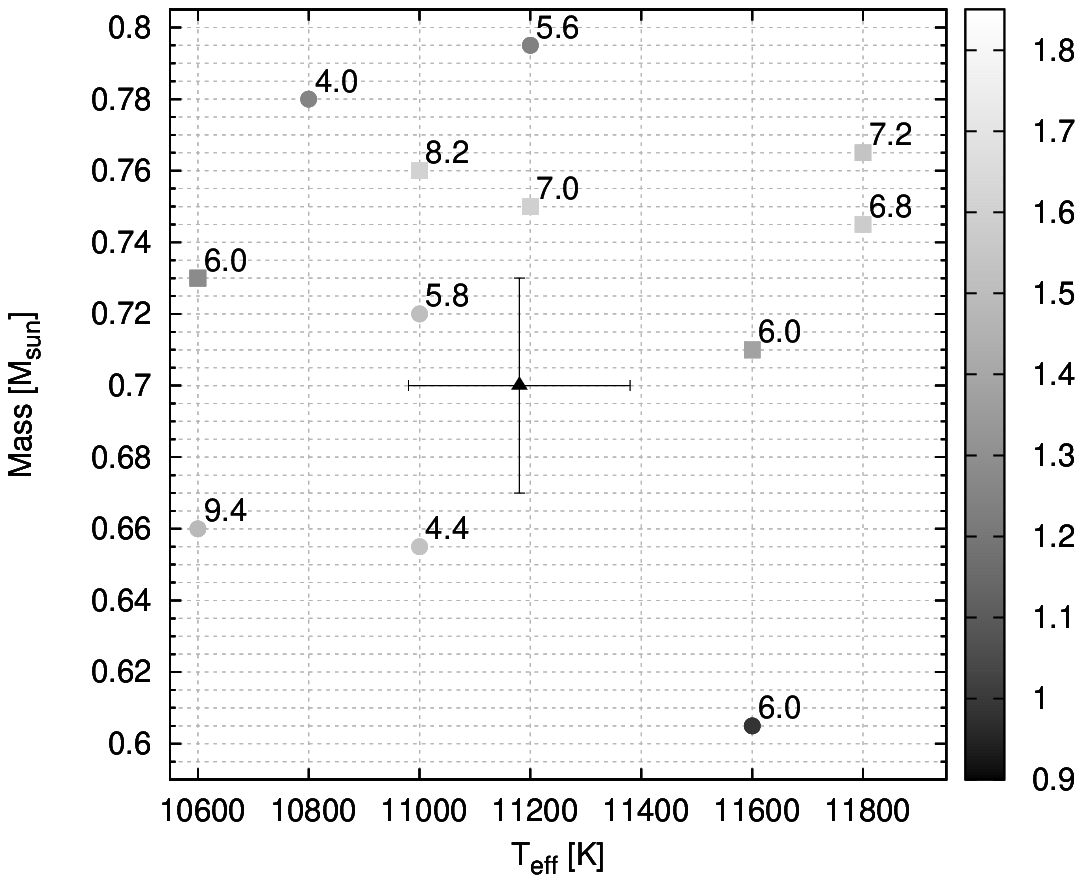}
\caption{ Plots of the models that best fit the data for six (left panel) or seven (right panel) modes 
in the $T_{\rmn{eff}}$\,--\,$M_*$ plane. Circles and squares denote the solutions obtained 
by including the 1191.7 or 1186\,s mode into the period list. Their $\sigma_{r.m.s.}$ values 
are colour-coded. The models' hydrogen layer masses (-log$M_{\rmn{H}}$), the spectroscopic 
$T_{\rmn{eff}}$ values, and their uncertainties are also indicated. The background grid 
corresponds to our model grid's step sizes.}
\label{fig:params1}
\end{figure*}

The $T_{\rmn{eff}}$ and $\rmn{log\,} g$ values of GD 154 determined by high signal-to-noise optical 
spectrophotometry are 11\,180\,K and 8.15\,dex, respectively \citep{bergeron2}. Considering that the 
external uncertainties are estimated to be hundreds of Kelvins and could achieve $\pm0.1$\,dex, we 
decided to cover a relatively large range in effective temperature and surface gravity. A DA white 
dwarf with $\rmn{log\,} g=$ 8.15 has a mass of $\sim$ 0.7\,$M_{\sun}$. In our grids, we searched for 
the best-fitting models between $\rmn{log\,} g\sim$ 8.0\,--\,8.3 (0.6\,--\,0.8\,$M_{\sun}$, see the 
tables of \citealt{bradley3}). \citet{bergeron04} estimated $\sim200$\,K and $\sim0.05$\,dex for the 
external errors of $T_{\rmn{eff}}$ and $\rmn{log\,} g$, respectively. According to these values, our 
grid covers $\pm3\sigma$ range in $T_{\rmn{eff}}$ and $M_*$. 

\citet{a2} suggested a very thin hydrogen layer for GD 154 with the mass of 
$2(\pm1)\,\times\,10^{-10}\,M_*$. Taking their result into account, we allowed for hydrogen layer
masses between $10^{-4}$ and $10^{-11}$\,$M_*$ in the grids.

We also used the database of ZZ Ceti periods (dipole and quadrupole) calculated from fully 
evolutionary models \citep{romero1}. The period values were derived by models with consistent 
chemical profiles from the core to the surface. The authors allowed the evolution of the stars 
from the zero-age main sequence in the calculations of these profiles. More details on the code, 
the input physics and a number of examples of its asteroseismological applications can be found 
in \citet{althaus1} and \citet{romero2}.

\subsection{Parameters of the best-fitting models}

We used two slightly different sets of periods for the asteroseismological investigations of 
the star. Our frequency analyses and tests show that the periods given in Table~\ref{tabl:final} 
describe the light variations of GD 154 in the 2006 observational season well. Accordingly, the 
values used in our first run are: {\it402.6, 1088.5, 1131.8, 1160.7, 1191.7} and {\it1238.2\,s}, 
respectively. The second run differs from this in one period only: we replaced the 1191.7\,s mode 
with {\it1186\,s}, the one with the second largest amplitude. During the 1991 WET observations, the 
period of the dominant mode was 1186.5\,s and was found to be the $m=0$ component of an (asymmetric) 
triplet \citep{a2}. As Tables~\ref{tabl:w2nd4th} and \ref{tabl:final} show in our dataset the 
1191.7\,s period dominates. This one is still part of a triplet structure and could be regarded
as the $m=-1$ peak of the 1186\,s mode. We used two different sets of periods
because of this ambiguity. This way, we could examine the effect of slightly different periods
on the best-fitting models' parameters.

Even though we doubled the number of known period values applied in the star's seismic investigations 
compared to previous research, GD 154 is still not a pulsator rich in known modes. This means that we find several models 
with low $\sigma_{r.m.s.}$ values as the result of the fitting procedure. Therefore, we applied 
further constraints during the model selection: assuming better visibility of $l=1$ modes, we 
selected the models with at least three $l=1$ solutions to the observed periods. As an additional 
criterion, we considered the ones which give $l=1$ value for the dominant (1186 or 1191.7\,s) mode.
However, as the average period spacing of the long-period modes is low (below 40\,s), we 
assumed that the observed modes are not solely $l=1$ ones, because this would require a much higher 
stellar mass for GD 154 than the spectroscopic value.    

\subsubsection{Stellar parameters}

The left panel of Fig.~\ref{fig:params1} shows the $6+6$ best-fitting models in the 
$T_{\rmn{eff}}$\,--\,$M_*$ plane for both period lists. The numbers indicate the models' 
hydrogen layer masses and we also denoted the spectroscopic solution given by \citet{bergeron2} 
with it's uncertainties. As it can be seen, we find the best-matching (three) models for 
the spectroscopic mass using the second period list. Table~\ref{tabl:params1} (rows 1-3) 
summarize the parameters of these models. In two cases ($M_*=0.71$ and $0.73$\,$M_{\sun}$) 
the mass of the hydrogen layer is around $10^{-6}$\,$M_*$, while the model star with 
$0.705$\,$M_{\sun}$ has $M_{\rmn{H}}=1.6*10^{-8}$\,$M_*$, which means a considerably thinner
 layer. The best-matching model to the spectroscopic parameters is the one with 
$M_*=0.73$\,$M_{\sun}$ and $T_{\rmn{eff}}=11\,200$\,K. In this case the mass of the 
hydrogen layer is $M_{\rmn{H}}=6.3*10^{-7}$\,$M_*$.

\begin{table*}
\caption{Parameters of the selected models using different core profiles and slightly 
different period values. The observed $T_{\rmn{eff}}$ and $M_*$ values \citep{bergeron2} 
are are given in the last row.}
\label{tabl:params1}
\begin{center}
\begin{tabular}[!ht]{p{1mm}cccccccccccccc}
\hline
No. & $T_{\rmn{eff}}$ & $M_*$ & -log$M_{\rmn{He}}$ & -log$M_{\rmn{H}}$ & $X_{\rmn{O}}$ & $X_{\rmn{fm}}$ & \multicolumn{7}{c}{Period values in seconds} & $\sigma_{r.m.s.}$\\
 & (K)             & ($M_{\sun}$) &               &                   &               &                & \multicolumn{7}{c}{($l, k$)} & (s)\\
\hline
1 & 11\,600 & 0.710 &  2.0 & 6.0 & 0.8 & 0.2 & 402.0 & 1088.9 & 1132.6 & 1159.2 & 1184.9 & 1238.4 & & 0.86\\
  &         &       &      &     &     &     & (1,7) & (1,23) & (2,42) & (2,43) & (1,25) & (2,46) & & \\
2 & 10\,800 & 0.705 &  2.0 & 7.8 & 0.6 & 0.4 & 402.9 & 1087.6 & 1131.5 & 1162.0 & 1188.1 & 1238.3 & & 1.09\\
  &         &       &      &     &     &     & (2,11) & (1,19) & (2,35) & (2,36) & (1,21) & (1,22) & & \\
3 & 11\,200 & 0.730 &  2.0 & 6.2 & 0.6 & 0.4 & 402.0 & 1090.8 & 1131.2 & 1161.3 & 1185.8 & 1239.2 & & 1.13\\
  &         &       &      &     &     &     & (1,6) & (1,22) & (2,40) & (2,41) & (1,24) & (2,44) & & \\
4 & 11\,000 & 0.720 &  2.0 & 5.8 & 0.8 & 0.5 & 400.4 & 1088.5 & 1132.2 & 1159.5 & 1194.2 & 1238.2 & 1273.4 & 1.51\\
  &         &       &      &     &     &     & (1,6) & (1,22) & (2,40) & (2,41) & (1,24) & (1,25) & (2,45) & \\
5 & 11\,600 & 0.710 &  2.0 & 6.0 & 0.8 & 0.2 & 402.0 & 1088.9 & 1132.6 & 1159.2 & 1184.9 & 1238.4 & 1268.5 & 1.38\\
  &         &       &      &     &     &     & (1,7) & (1,23) & (2,42) & (2,43) & (1,25) & (2,46) & (1,27) & \\
6 & 11\,200 & 0.675 &  2.0 & 5.2 & 0.73 & 0.54 & 402.4 & 1089.2 & 1133.6 & 1158.1 & 1191.9 & 1240.1 & & 1.52\\
  &         &       &      &     &      &      & (2,12) & (1,22) & (2,41) & (2,42) & (1,24) & (1,25) & & \\
7 & 11\,400 & 0.675 &  2.5 & 5.2 & 0.73 & 0.54 & 402.2 & 1086.5 & 1132.7 & 1160.5 & 1194.5 & 1239.4 & & 1.55\\
  &         &       &      &     &      &      & (1,6) & (1,22) & (2,41) & (2,42) & (1,24) & (2,45) & & \\
8 & 11\,000 & 0.710 &  2.0 & 5.6 & 0.73 & 0.54 & 398.9 & 1085.6 & 1131.4 & 1160.7 & 1188.0 & 1236.8 & & 2.16\\
  &         &       &      &     &      &      & (1,6) & (1,22) & (2,41) & (2,42) & (1,24) & (1,25) & & \\
9 & 11\,200 & 0.700 &  2.0 & 4.2 & 0.73 & 0.54 & 398.9 & 1085.9 & 1132.8 & 1159.9 & 1192.7 & 1236.0 & 1269.9 & 2.07\\
  &         &       &      &     &      &      & (1,7) & (2,43) & (2,45) & (2,46) & (1,27) & (1,28) & (1,29) & \\
10 & 11\,000 & 0.710 & 2.0 & 5.6 & 0.73 & 0.54 & 398.9 & 1085.6 & 1131.4 & 1160.7 & 1188.0 & 1236.8 & 1267.7 & 2.45\\
  &         &       &      &     &      &      & (1,6) & (1,22) & (2,41) & (2,42) & (1,24) & (1,25) & (2,46) & \\
11 & 11\,241 & 0.705 &     & 4.445 &    &      & 399.3 & 1082.6 & 1133.9 & 1160.0 & 1192.3 & 1239.8 & & 2.97\\
  &         &       &      &     &      &      & (1,7) & (2,43) & (2,45) & (2,46) & (1,27) & (1,28) & & \\
12 & 11\,639 & 0.705 &     & 9.339 &    &      & 404.3 & 1085.3 & 1130.7 & 1165.7 & 1182.6 & 1237.9 & & 2.92\\
  &         &       &      &     &      &      & (1,5) & (1,20) & (1,21) & (2,38) & (1,22) & (1,23) & & \\
\multicolumn{3}{l}{Observations:} & & & & & & & & & & & &\\
 & 11\,180 & 0.70 &         &       &    &      & 402.6 & 1088.5 & 1131.8 & 1160.7 & 1191.7 & 1238.2 & 1271.5 & \\
 &         &     &         &       &    &      &       &        &        &        & 1186.0 &        &        & \\
\hline
\end{tabular}
\end{center}
\end{table*}

\begin{figure}
\begin{center}
 \includegraphics[width=8.7cm]{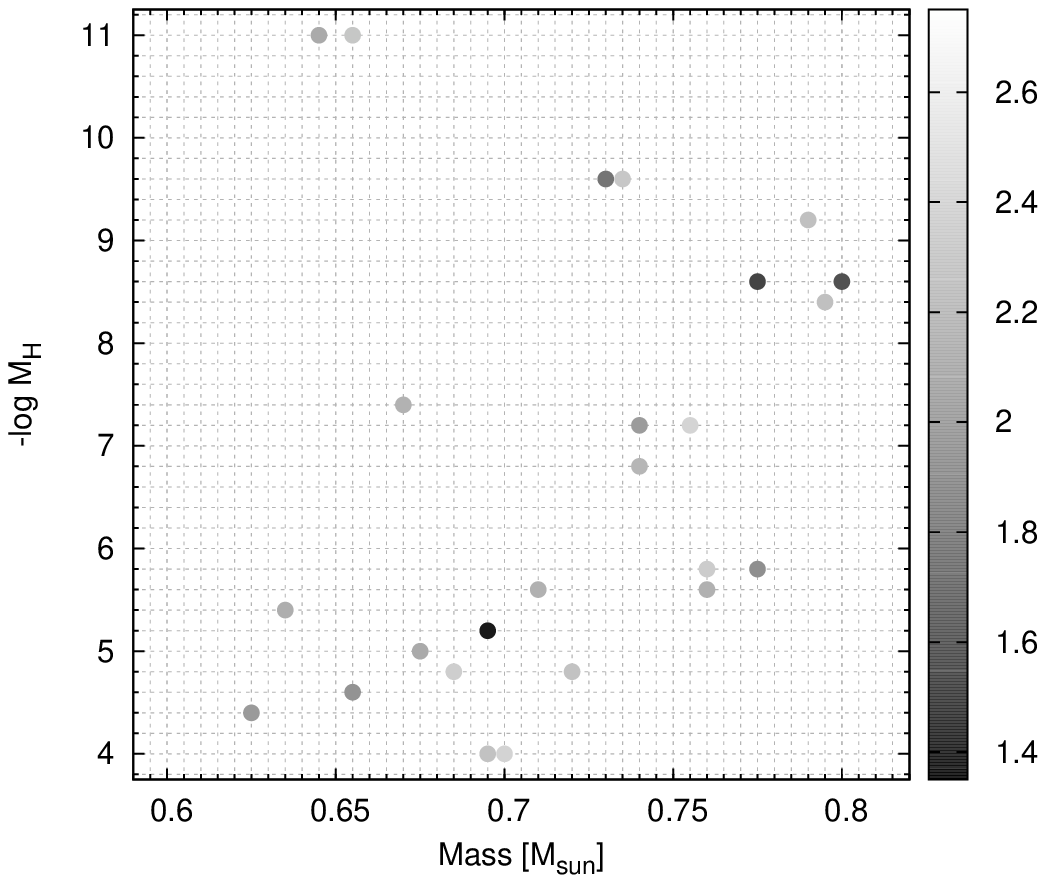}
\end{center}
 \caption{Models in the $M_*$\,--\,$M_{\rmn{H}}$ plane with $\sigma_{r.m.s.}<2.5$\,s and 
Salaris et al.'s core profiles. The six-period light curve solution with the 1186\,s mode 
was used. The models' $\sigma_{r.m.s.}$ values are colour-coded. The background grid corresponds 
to our model grid's step sizes. The figure shows the trend that more massive stellar models 
have thinner hydrogen layer.}
\label{fig:params3}
\end{figure}

When we investigate a pulsator showing amplitude variations, it is worth checking  
if we can add further modes to our period lists observed in a different season. We found 
one mode, also presented by \citet{a14}: the one at 1271.5\,s. Then we have four period lists: 
two with 6-6 and two with 7-7 periods. In the latter case we selected the models with at 
least four $l=1$ modes instead of three. As the right panel of Fig.~\ref{fig:params1} 
shows, the parameter space occupied by the best fitting seven-period models differ only 
slightly from the six-period ones. Using seven periods most of the models have 
$M_*>0.7\,M_{\sun}$. Considering the hydrogen layer 
masses, the average values are $\sim10^{-7}$ and $4*10^{-7}\,M_*$ for the six- and 
seven-period solutions, respectively. We can recognize in Fig.~\ref{fig:params1} that there 
are common models, the ones with $T_{\rmn{eff}}=10\,600$ and $11\,600$\,K and 
$M_{\rmn{H}}=10^{-9.4}$ and $10^{-6}\,M_*$. Our best-matching models using seven periods are 
the $0.720$ and $0.710$\,$M_{\sun}$ ones with $T_{\rmn{eff}}=11\,000$\,K and $11\,600$\,K. 
They have $M_{\rmn{H}}=1.6*10^{-6}$ and $10^{-6}\,M_*$ (Table~\ref{tabl:params1}, rows 4-5). 
As can be seen, models No. 1 and 5 have the same parameters, this is one of the common points.  

\begin{table*}
\caption{Physical parameters and mode-identification results presented by different authors. 
They selected the model solutions using the period values determeined by the 1991 WET campaign 
\citep{a2}. The atmospheric parameters and periods obtained by observations are denoted by 
asterisks.}
\label{tabl:others}
\begin{center}
\begin{tabular}[!ht]{lccclllc}
\hline
$T_{\rmn{eff}}$ (K) & $M_*$ ($M_{\sun}$) & -log\,$M_{\rmn{He}}$ & -log\,$M_{\rmn{H}}$ & \multicolumn{3}{c}{Period values in seconds} & Ref. \\
\hline
11\,180$^*$ & 0.7$^*$  &     & 9.7 & 402.6$^*$(1)   & 1088.6$^*$(2)   & 1186.5$^*$(1) & \citet{a2}\\
11\,200 & 0.68 & 2.0 & 7.5 & 398.2(1,5) & 1088.5(1,19) & 1186.9(1,21) & \citet{castanheira1}\\
10\,800 & 0.73 & 2.5 & 9.5 & 396.9(1,4) & 1088.5(1,18) & 1186.5(1,20) & \citet{castanheira1}\\
11\,574 & 0.705 & 2.1 & 9.3 & 405.0(1,5) & 1088.9(1,20) & 1186.6(1,22) & \citet{romero2}\\
\hline
\end{tabular}
\end{center}
\end{table*}

Considering the solutions with Salaris et al.'s core profiles and within the $1\sigma$ 
limit in $M_*$, the hydrogen layer masses are between $10^{-4}$ and $10^{-6}$\,$M_*$ in 
most cases and the $M_{\rmn{He}}=10^{-2}\,M_*$ value is preferred. Rows 6-10 of 
Table~\ref{tabl:params1} show some of their parameters; they are selected on the basis of 
having $T_{\rmn{eff}}$ close to the spectroscopic value. These models have 
$M_{\rmn{H}}=6.3*10^{-5}$, $2.5*10^{-6}$ or $6.3*10^{-6}\,M_*$.

\begin{figure*}
\begin{center}
 \includegraphics[width=17.5cm]{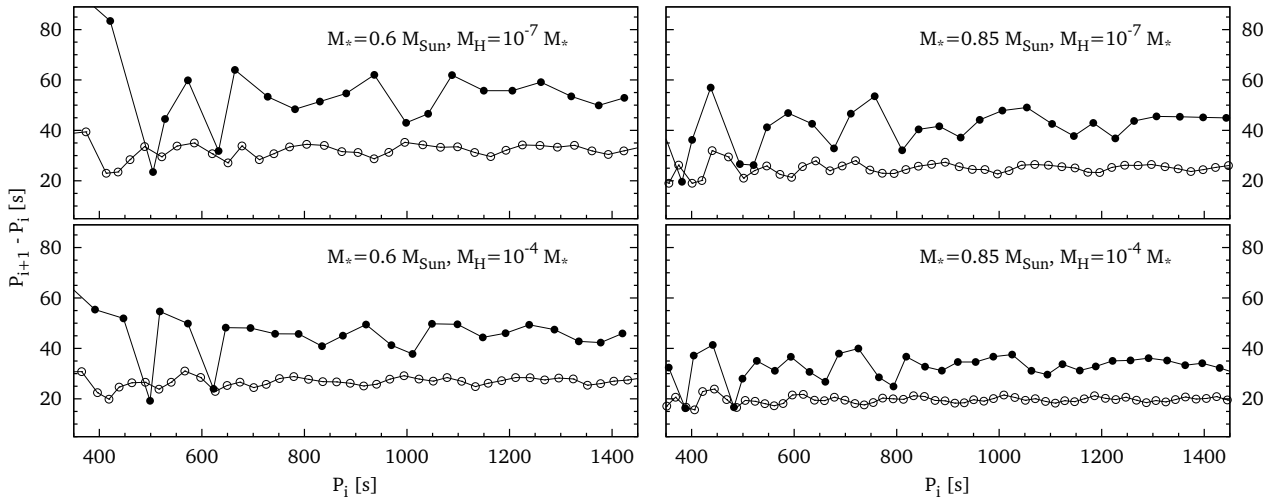}
\end{center}
 \caption{Period spacing diagrams of model stars with $T_{\rmn{eff}}=11\,200$\,K, $M_{\rmn{He}}=10^{-2}$\,$M_*$ and 
 $50/50$ C/O fraction core. Filled and open circles denote the $l=1$ and 2 modes, respectively. 
 The panels show the influence of the stellar and hydrogen layer mass variations on the
 period spectrum.}
\label{fig:spacing2}
\end{figure*}

Investigating the parameters of the best-fitting models, we found a trend that more massive 
stellar models have a thinner hydrogen layer. An example for the phenomenon can be seen in 
Fig.~\ref{fig:params3}, where we plotted the best-fitting models with $\sigma_{r.m.s.}<2.5$\,s 
in the $M_*$\,--\,$M_{\rmn{H}}$ plane. The explanation of this phenomenon is that generally 
for thinner hydrogen layers, the mode trapping cycle is longer and the average period spacings 
between the consecutive overtone modes are also larger (i.e. there are larger differences between period
spacing minima and maxima). This can be partially offset by the shorter 
average period spacing of a higher mass star \citep{bradley3}. 
The panels of Fig.~\ref{fig:spacing2} demonstrate
the influence of the varying stellar parameters on the period spectrum of a model star. The stellar mass and the mass
of the hydrogen layer has a great effect on the observed periods. We
can find a similar period structure for the lower stellar mass --
higher hydrogen layer mass and higher stellar mass -- thinner hydrogen
layer models, and this results in the observed trend seen in Fig.~\ref{fig:params3}. 

Using the coarse grid of \citet{romero1} we obtained the best-matching model to 
spectroscopy applying the first period set. It has $T_{\rmn{eff}}=11\,241$\,K, 
$M_*=0.705$\,$M_{\sun}$ and $M_{\rmn{H}}=3.6*10^{-5}$\,$M_*$ (Table~\ref{tabl:params1}, 
row 11). The second best-matching model belongs to the second period list and has higher 
effective temperature ($11\,639$\,K) but the same mass (Table~\ref{tabl:params1}, row 12). 
The largest difference is in the mass of the hydrogen layer, which is significantly lower, 
only $4.6*10^{-10}$\,$M_*$. Considering all the solutions within the $\sigma_{r.m.s.}<3$ 
limit, the mass of the hydrogen layer is between $3.5$ and $7.6*10^{-5}$\,$M_*$, from which 
the latter one is the only exception.

Knowing the luminosities of the selected models, we estimated the star's asteroseismological 
distance by calculating the distance modulus \citep{bradley5}. The log($L/L_{\sun}$) 
values of the models in Table~\ref{tabl:params1} are between $-2.7$ and $-2.8$. The seismological 
distances calculated from the ten different models of Table~\ref{tabl:params1} 
are between $41.5$ and $46.4$\,pc. The average distance and parallax value is $44.2$\,pc and $22.7$\,mas, 
respectively. This result is close to the $45.4$\,pc value derived by spectroscopic observations 
of \citet{lajoie1}, which shows that the pulsation analysis supports the spectroscopic result 
on the distance parameter.

For the sake of completeness, we summarized the results obtained by previous asteroseismological 
investigations of GD 154 in Table~\ref{tabl:others}.

\subsubsection{Mode identification}

As it was already mentioned, we placed a constraint only on the $l$ value of the dominant 
mode: we assumed that it is $l=1$. All of the other modes were allowed to be $l=1$ or $2$. 
Considering the $l$ and $k$ values of the selected models in Table~\ref{tabl:params1}, we 
obtained the same $l$ only in the case of the 1160\,s mode. This result suggests that it 
may be an $l=2$. None of the selected models gives $l=1$ solutions for all of the periods. 
We cannot uniquely assign an $l$ value for the other modes, however, the 402, 1088 and 1131\,s 
modes' $l$ values are 1, 1 and 2, respectively, in the vast majority of cases. Assuming that 
the 1131 and 1160\,s modes are $l=2$, they represent consecutive overtones.

The models selected by \citet{castanheira1} and \citet{romero2} (see Table~\ref{tabl:others}) 
give $l=1$ solutions for the 402, 1088 and 1186\,s modes as well.

\subsubsection{Mode trapping and the mass of the hydrogen layer}

\citet{a2} assumed that mode trapping -- as an efficient mode selection mechanism -- could 
be responsible for the small number of observed modes in GD 154 and determined a very low 
mass for the hydrogen layer ($\sim10^{-10}\,M_*$). As Table~\ref{tabl:others} shows, the 
results of the previous asteroseismological investigations also supported the presence of a very 
thin hydrogen layer. However, considering our selected models in Table~\ref{tabl:params1}, we 
found that the masses of hydrogen layers were higher than expected. Our model solutions have 
$M_{\rmn{H}}$ values between $6.3*10^{-5}$ and $6.3*10^{-7}\,M_*$ (with an average value of 
$3.9*10^{-6}$) except for two cases, when $M_{\rmn{H}}=1.6*10^{-8}$ and $4.6*10^{-10}\,M_*$. 
However, these latter models are below or above the $1\sigma$ limit in effective temperature. 
Our results show that although GD 154 may have thinner hydrogen layer than the maximum allowed 
($M_{\rmn{H}}\approx10^{-4}$), no extremely thin layer is necessary to explain the sequence 
of the observed periods.   

The question is raised: why the other authors obtained lower mass values for the hydrogen layer. 
The answer may be the number of modes used in the period fits. Both \citet{castanheira1} and 
\citet{romero2} worked with the three periods determined by \citet{a2}. This fact obviously 
strongly influenced the outcome of the fit. The results of \citet{a2} have not been obtained
by performing period fits with a model grid. Because of the small number of known
modes they needed to impose further constraints. As mode trapping means an efficient mode selection 
mechanism, they assumed that the observed modes are trapped in the outer hydrogen layer and 
derived the $M_{\rmn{H}}\approx10^{-10}\,M_*$ value. In our case, the increased number of observed 
modes is sufficient for the asteroseismic investigations using model grids, including searching for
trapped modes among the observed ones.   

\begin{figure}
\begin{center}
 \includegraphics[width=8.6cm]{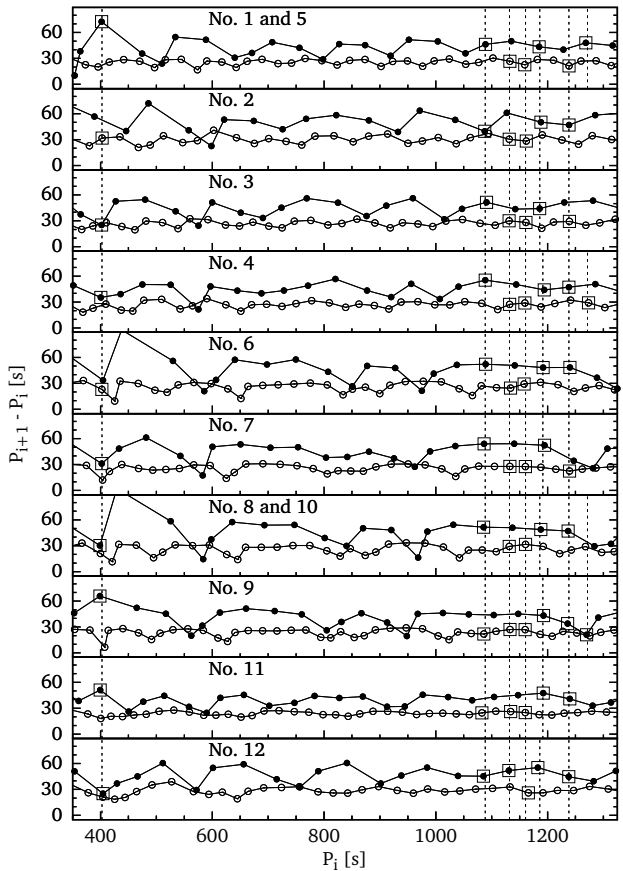}
\end{center}
 \caption{Period spacing diagrams of the ten different models of Table~\ref{tabl:params1}. 
Filled and open circles denote the $l=1$ and 2 modes, respectively. Vertical dashed lines mark 
the observed period values. We denoted the periods analogous to the observed ones in the 
given model with open squares. The models' numbers corresponding to Table~\ref{tabl:params1} 
are also indicated in the panels.}
\label{fig:spacing}
\end{figure}

We constructed the (forward) period spacing diagrams for our selected models (Fig.~\ref{fig:spacing}). 
Minima in the diagrams denote departures from the uniform period spacings caused by mode trapping 
(see e.g. \citealt{bradley4}). Fig.~\ref{fig:spacing} shows that the long-period modes do not show 
tendency to occur around minima, so our model selection does not support the idea of mode trapping 
at long periods as a mode selection mechanism. 
The best candidate for a trapped mode is the 1238\,s one according to 
Fig.~\ref{fig:spacing}. However, the situation is different in the case of the only short-period 
mode. In most cases we can find it near or at a period-spacing minimum. This suggests that the 
402\,s mode could be a trapped mode.     

\section{Summary and conclusions}

Our comparative period search confirmed the frequencies found previously, except the one 
at 787\,$\mu$Hz given by \citet{a14}. Additionally, we localized modes at 807.62 and 861.56\,$\mu$Hz 
values that have never been reported before. We confirmed by test investigations, that 
six modes can be considered as independent normal modes of pulsation.

Four doublets around the largest amplitude modes were directly found. The dominant mode 
of our whole data set (839.14\,$\mu$Hz) and the dominant mode observed in two previous pulsation 
stages (843.15\,$\mu$Hz) are members of the same rotational triplet. This latest member of the triplet 
became dominant at the end of our observing run, when GD 154 also presented a monoperiodic pulsational stage.

We localized the second harmonic of the mother frequency at 2517.48\,$\mu$Hz, near the high frequency 
normal mode at 2484.14\,$\mu$Hz, and some linear combinations too. However, no subharmonics reported by previous 
observations were found over our 
whole observational season. Characteristic features are localized in the light curves partly suggesting 
an effect of convection to the pulsation and reminding us of the chaotic behaviour of stellar pulsation.

Comparative analyses of subsets revealed and test investigations confirmed a remarkable intrinsic 
amplitude change of frequencies at
839.14 and 861.56\,$\mu$Hz, although part of it can be caused by unresolved rotational triplets.

With our new frequencies we have doubled the number of period values applied in the star's previous seismic 
investigations, 
which allows a more detailed study of GD 154. We found models with effective temperatures
and masses within the $1\sigma$ limit of the spectroscopic values ($\approx11\,000 - 11\,400$\,K 
and $0.68 - 0.73\,M_*$) that fit the observed periods 
well and also give $l=1$ solutions for at least half of the modes. The best-fitting models have 
hydrogen layer masses between $6.3*10^{-5}$ and $6.3*10^{-7}\,M_*$,
which suggests orders of magnitudes thicker layer than previously published. The explanation of this
difference may be the number of modes used for the seismic studies and that we did not assume
that our observed modes were trapped ones. Considering the mass of the
helium layer, our results also supports the $10^{-2}\,M_{\rmn{He}}$ `canonical' value.

The known luminosities of our selected models allowed us to determine the seismic distance of GD 154. In
agreement with other authors' finding, the average value calculated by our selected models is 
$44$\,pc. We also investigated the possibility of mode trapping, constructing period spacing diagrams. 
Our results do not
support the idea of mode trapping at long periods as a mode selection
mechanism, and suggest that the 
short-period mode may be a real trapped mode. This result shows that we do not have to presume that 
all the observed modes are trapped ones. 

Both the new frequency content, the altering pulsation between the multi- and monoperiodic stage and also
the need for more constraints for modeling express 
a requirement for further investigation of the complex behaviour of GD 154. Regular monitoring of the star
would be necessary not only to detect new pulsation modes for asteroseismic investigations but to follow-up
more mono- and multiperiodic stages. This might allow us to determine whether this altering between the two
pulsation stages has any regular behaviour. 

\section*{Acknowledgments}

The authors thank Agn\`es Bischoff-Kim for providing her version of the
WDEC and the \textsc{fitper} program. The authors also acknowledge the contribution of P. I. P\'apics,
E. Bokor, Gy. Kerekes, A. M\'ar and N. Sztank\'o to the observations of GD 154.

\bsp

\label{lastpage}

\end{document}